\documentclass[aps,prl,twocolumn,superscriptaddress]{revtex4}

\usepackage{graphicx}

\begin{document}

\title{Spin correlations in the electron-doped high-transition-temperature \\
superconductor {Nd$_{2-x}$Ce$_x$CuO$_{4\pm\delta}$}}

\author{E.M.~Motoyama} \affiliation{Department of Physics, Stanford
University, Stanford, California 94305, USA}

\author{G.~Yu} \affiliation{Department of Physics, Stanford University,
Stanford, California 94305, USA}

\author{I.M.~Vishik} \affiliation{Department of Physics, Stanford
University, Stanford, California 94305, USA}

\author{O.P.~Vajk} \affiliation{NIST Center for Neutron Research, National
Institute of Standards and Technology, Gaithersburg, Maryland 20899, USA}

\author{P.K.~Mang} \affiliation{Department of Applied Physics, Stanford
University, Stanford, California 94305, USA}

\author{M.~Greven} \email[]{greven@stanford.edu} \affiliation{Department of
Applied Physics, Stanford University, Stanford, California 94305, USA}
\affiliation{Stanford Synchrotron Radiation Laboratory, Stanford, California
94309, USA}

\maketitle

\textbf{
High-transition-temperature (high-$T_\mathrm{c}$) superconductivity develops
near antiferromagnetic phases, and it is possible that magnetic excitations
contribute to the superconducting pairing mechanism.  To assess the role of
antiferromagnetism, it is essential to understand the doping and temperature
dependence of the two-dimensional antiferromagnetic spin correlations.  The
phase diagram is asymmetric with respect to electron and hole doping, and
for the comparatively less-studied electron-doped materials, the
antiferromagnetic phase extends much further with doping
\cite{KeimerPRB92,MatsudaPRB92} and appears to overlap with the
superconducting phase. The archetypical electron-doped compound
Nd$_{2-x}$Ce$_x$CuO$_{4\pm\delta}$ (NCCO) shows bulk superconductivity above
\emph{x} $\approx$ 0.13 \cite{TakagiPRL89,UefujiPhysicaC01}, while evidence
for antiferromagnetic order has been found up to \emph{x} $\approx$ 0.17
\cite{MatsudaPRB92,UefujiPhysicaC02,MangPRL04}. Here we report inelastic
magnetic neutron-scattering measurements that point to the distinct
possibility that genuine long-range antiferromagnetism and superconductivity
do not coexist. The data reveal a magnetic quantum critical point where
superconductivity first appears, consistent with an exotic quantum phase
transition between the two phases \cite{SenthilScience04}. We also
demonstrate that the pseudogap phenomenon in the electron-doped materials,
which is associated with pronounced charge anomalies
\cite{ArmitagePRL02,OnosePRB04,MatsuiPRL05a,ZimmersEPL05}, arises from a
build-up of spin correlations, in agreement with recent theoretical
proposals \cite{KyungPRL04,MarkiewiczPRB04}.
}

In their as-grown state, the electron-doped materials exhibit
antiferromagnetic (AF) order throughout the accessible doping range, and an
oxygen reduction treatment is required to induce superconductivity
\cite{TakagiPRL89}. Previous inelastic neutron scattering experiments on
as-grown, non-superconducting (non-SC) NCCO demonstrated \cite{MangPRL04}
that the two-dimensional (2D) spin correlation length $\xi$ observed above
the N\'eel temperature is exponentially dependent on inverse temperature:
\begin{equation}
\xi(x,T) = A(x) \exp(2\pi\rho_s(x)/T) \label{xiexp}
\end{equation}
for cerium concentrations ranging from zero up to the solubility limit of
$x\approx 0.18$. This behaviour indicates the existence of an underlying
ground state with long-range 2D AF order. Owing to weak spin-space
anisotropies and three-dimensional couplings, NCCO exhibits
three-dimensional AF order at a non-zero N\'eel temperature, as observed in
the elastic scattering channel \cite{MatsudaPRB92,MangPRL04}. The spin
stiffness $\rho_s(x)$ decreases monotonically with increasing electron
concentration, with $\rho_s(0.18)/\rho_s(0) \approx 25\%$ \cite{MangPRL04},
and the doping dependence of $\rho_s(x)$ and of the amplitude $A(x)$ is
remarkably close to that for the randomly-diluted spin-one-half
square-lattice Heisenberg antiferromagnet La$_2$Cu$_{1-z}$(Zn,Mg)$_z$O$_4$
\cite{VajkScience02}.

\begin{figure}[b] \includegraphics[width=3.0in]{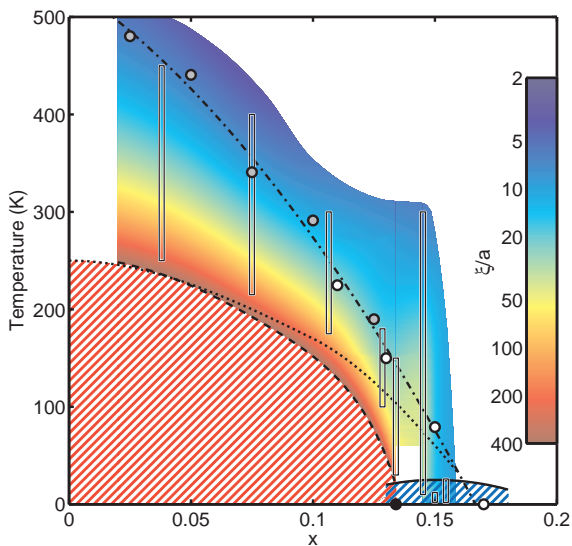}
\caption{\label{phasediagram} \textbf{The temperature-doping phase diagram
for oxygen-reduced NCCO.} The red and blue hashed areas indicate long-range
AF order and superconductivity, respectively. The black circle at zero
temperature indicates the approximate location of a magnetic quantum phase
transition. The instantaneous 2D spin-correlation length $\xi(x,T)$ in the
CuO$_2$ sheets was measured at the doping levels and over the temperature
ranges indicated by the vertical bars.  The colour scale shows $\xi$, in
units of the planar lattice constant $a$, interpolated and extrapolated from
the measured values. The N\'{e}el temperature $T_\mathrm{N}$ is shown as the
dotted curve, while the dashed curve is the extrapolated contour of $\xi/a =
400$. The measurement of $T_\mathrm{N}$ is contaminated by remnants of the
as-grown state of NCCO, so that the true AF phase extends only to
$x_\mathrm{AF} \approx 0.13$, close to where superconductivity first
appears. This is established from the fact that $\xi$ diverges exponentially
upon cooling for non-superconducting compositions at lower electron
concentrations, while it remains finite in superconducting samples. The
small remaining overlap indicated in the figure may be caused by cerium and
oxygen inhomogeneities.  The grey and white circles indicate optical
conductivity measurements of of the pseudogap temperature $T^*$ on NCCO
crystals \cite{OnosePRB04} and Pr$_{2-x}$Ce$_x$CuO$_{4\pm\delta}$ thin films
\cite{ZimmersEPL05}, respectively. The dot-dashed curve is a guide to the
eye.}
\end{figure}

\begin{figure}[!b] \includegraphics[width=3.0in]{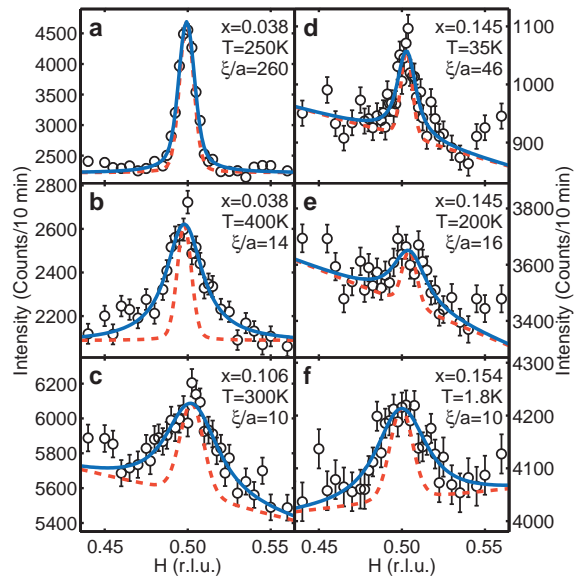}
\caption{\label{scans} \textbf{Representative two-axis scans used to measure
the spin correlation length.} The scans are along $(h, h)$ about the 2D AF
zone centre $(1/2,1/2)$ and are fitted (solid blue curves) to a 2D
Lorentzian convoluted with the calculated instrumental resolution (dashed
red curves). Shown are data at {\bf\sffamily a}, $T = 250\,\mathrm{K}
\approx T_\mathrm{N}$ and {\bf\sffamily b}, $T = 400\,\mathrm{K}$ for
$x=0.038$; {\bf\sffamily c}, $T = 300\,\mathrm{K}$ for $x=0.106$;
{\bf\sffamily d}, $T = 35\,\mathrm{K}$ and {\bf\sffamily e}, $T =
200\,\mathrm{K}$ for $x=0.145$; and {\bf\sffamily f}, $T = 1.8\,\mathrm{K}$
for $x=0.154$. Wavevectors are represented as $(h, k, l)$ in reciprocal
lattice units (r.l.u.), where $Q = (2\pi h/a, 2\pi k/a, 2\pi l/c)$ is the
momentum transfer, and $a$ and $c$ are the lattice constants of the
tetragonal system (space group $I4/mmm$; for $x=0.038$, for example, the
room-temperature lattice constants are $a = 3.93\,\textrm{\AA}$ and $c =
12.09\,\textrm{\AA}$). Vertical error bars represent uncertainties
($1\sigma$) assuming Poisson statistics. The measurements were performed in
two-axis mode on the BT2 and BT9 triple-axis spectrometers at the NIST
Center for Neutron Research. The incident neutron energy was $E_\mathrm{i} =
14.7\,\mathrm{meV}$. In previous experiments on
La$_2$Cu$_{1-z}$(Zn,Mg)$_z$O$_4$ \cite{VajkScience02} and as-grown NCCO
\cite{MangPRL04}, this energy proved to be sufficiently large in the
temperature region $T_\mathrm{N} < T < 2T_\mathrm{N}$ to reliably extract
the instantaneous structure factor $S(Q)$. The collimations were:
{\bf\sffamily a}, {\bf\sffamily b}, 40$'$-23$'$-sample-20$'$; {\bf\sffamily
c}, 60$'$-40$'$-sample-40$'$; {\bf\sffamily d}, 40$'$-47$'$-sample-10$'$;
{\bf\sffamily e}, 40$'$-47$'$-sample-20$'$; and {\bf\sffamily f},
40$'$-47$'$-sample-40$'$. The NCCO crystals were grown in $4\,\mathrm{atm}$
of oxygen using the travelling-solvent floating-zone technique, and
subsequently annealed for 10 h at $970{^\circ}\mathrm{C}$ in argon, followed
by 20 h at $500{^\circ}\mathrm{C}$ in oxygen.  The sample masses range from
1 to $5\,\mathrm{g}$.  The oxygen reduction treatment, required for
superconductivity to appear, is a non-equilibrium process resulting in
unavoidable oxygen inhomogeneities. Cerium concentrations $x$ were
determined from inductively coupled plasma (ICP) spectroscopy, with typical
variation of ${\Delta}x \approx 0.005$ along the growth direction.
Superconductivity is observed from magnetic susceptibility measurements for
$x\ge0.134$.}
\end{figure}

\begin{figure}[b] \includegraphics[width=3.0in]{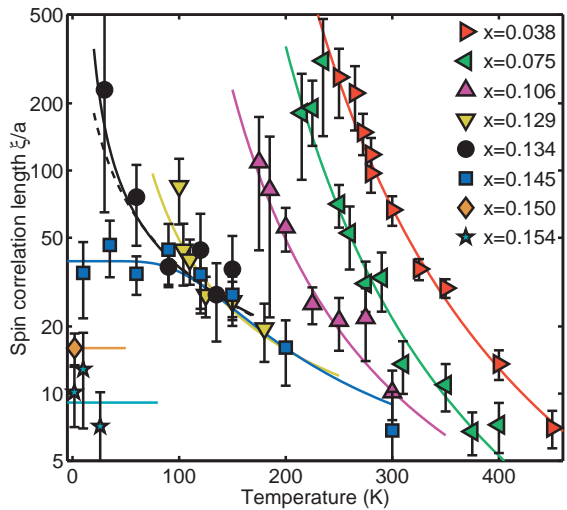}
\caption{\label{xixt} \textbf{The temperature dependence of the spin
correlation length at various cerium concentrations.} Vertical bars
represent uncertainties of $1\sigma$. The data for $x \le 0.134$ are fit to
equation~(\ref{xiexp}). The spin stiffness may already be zero for
$x=0.134$, because a fit to a simple power law $\xi \propto 1/T$ describes
the data equally well (dashed curve); power-law behaviour is expected at a
quantum critical point. For $x=0.145$ and higher, $\xi$ does not diverge,
but instead remains finite at low temperatures, demonstrating the absence of
genuine long-range AF order. The curves drawn for these latter data are
guides to the eye. Superconductivity is observed for $x\ge0.134$.}
\end{figure}

\begin{figure}[!b] \includegraphics[width=3.0in]{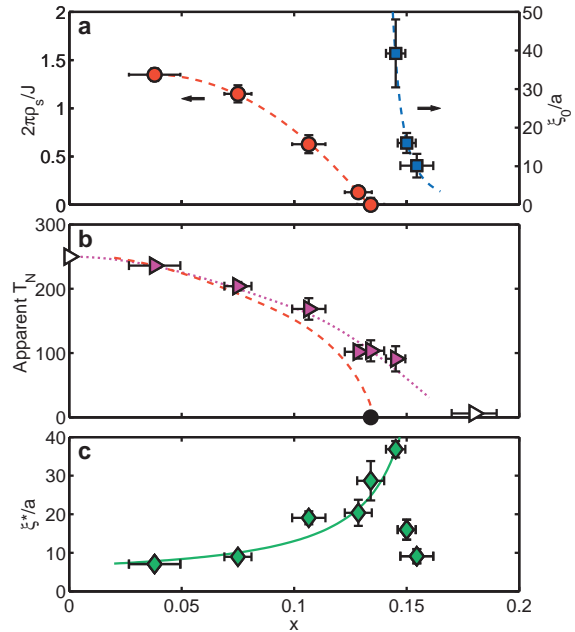}
\caption{\label{stiffness} \textbf{Spin stiffness, spin correlations at
low-temperature, apparent N\'eel temperature, and spin correlations along
$T^*$.} {\bf\sffamily a}, Doping dependence of the spin stiffness $\rho_s$
(plotted as $2\pi\rho_s/J$, where $J=125\,\mathrm{meV}$ is the AF
superexchange for the undoped Mott insulator Nd$_2$CuO$_4$
\cite{MatsudaPRB92,MangPRL04}) and of the low-temperature spin correlation
length $\xi_0$. Vertical error bars here represent uncertainties of
$1\sigma$. Horizontal error bars in all panels represent the measured range
of cerium concentration in each crystal. Dashed curves are guides to the
eye.  The spin stiffness decreases smoothly with doping and reaches zero in
an approximately linear fashion around $x_\mathrm{AF} \approx 0.134$.  The
ground state for $x<x_\mathrm{AF}$ has long-range AF order, whereas
long-range order is absent for $x>x_\mathrm{AF}$, as seen from the finite
values of $\xi_0$. The doping dependence of $\xi_0$ indicates a divergence
as the critical point is approached from the right. {\bf\sffamily b},
Apparent N\'{e}el temperature $T_\mathrm{N}$, as determined from elastic
scattering, as a function of doping.  The temperature dependence of the
measured order parameter (not shown) was modelled using a gaussian
distribution of $T_\mathrm{N}$, and the vertical bars indicate the
full-width at half-maximum (FWHM) of this distribution.  Measurements were
not performed on all samples; previous data \cite{MangPRL04} are indicated
by open symbols. The dotted and dashed curves are the same as in
Fig.~\ref{phasediagram}. {\bf\sffamily c}, The spin correlation length
$\xi^*$ measured at or extrapolated to the pseudogap temperature $T^*$.
Vertical bars represent uncertainties of $1\sigma$. Below optimal doping
($x<0.15$), $\xi^*$ is given by the single-particle thermal de Broglie
wavelength and increases as $\xi^* \propto 1/(x^* - x)$ (fitted curve).
However, this relationship breaks down near optimal doping, where $\xi^*$ is
found not to exceed the SC coherence length.}
\end{figure}

Magnetic inelastic neutron-scattering experiments in the SC phase have
become possible only in recent years \cite{YamadaPRL03,MotoyamaPRL06}. We
have carried out two-axis measurements of the spin correlations in eight
oxygen-reduced NCCO crystals in the cerium concentration range $0.038 \leq x
\leq 0.154$ (Fig.~\ref{phasediagram}). The data are fit to a 2D lorentzian,
$S(q_\mathrm{2D}) = S(0) / (1+q_\mathrm{2D}^2 \xi^2)$, convoluted with the
calculated instrumental resolution, where $q_\mathrm{2D}$ is the distance in
momentum space from the 2D AF zone center (Fig.~\ref{scans}). The non-SC
samples ($x\le0.129$) follow equation~(\ref{xiexp}), consistent with bulk AF
order in the ground state (Fig.~\ref{xixt}). Although this behaviour is
qualitatively the same as that for as-grown NCCO \cite{MangPRL04}, the spin
stiffness decreases much more rapidly with doping. The data for the SC
sample with $x=0.134$ are fit to equation~(\ref{xiexp}) with a small value
of the spin stiffness $\rho_s$, but are equally well described by the simple
power law $\xi \propto 1/T^{\nu_T}$ with exponent $\nu_T = 1.0(5)$.  The
power-law behaviour, indicated by the dashed curve in Fig.~\ref{xixt}, would
imply that $\rho_s$ is already zero and that the system is quantum critical
at this cerium concentration. Fig.~\ref{stiffness}a demonstrates that
$\rho_s$ approaches zero at $x_\mathrm{AF}=0.134(4)$ in an approximately
linear fashion. In a fundamental departure from the above behaviour, we find
that in the SC samples with $x\ge 0.145$, $\xi$ remains finite down to the
lowest temperatures. The low-temperature correlation length $\xi_0$ for
these samples increases as $x_\mathrm{AF}$ is approached from above
(Fig.~\ref{stiffness}a).

Previous elastic neutron measurements
\cite{MatsudaPRB92,UefujiPhysicaC01,MangPRL04} indicated that oxygen-reduced
NCCO exhibits AF order up to $x\approx0.17$. Such measurements in our
crystals indeed reveal N\'{e}el order.  However, our inelastic results
demonstrate that, contrary to previous belief
\cite{UefujiPhysicaC01,ZimmersEPL05,DaganPRL05}, the ground state of SC
samples exhibits only short-range spin correlations. Moreover, prior
inelastic neutron-scattering experiments clearly revealed a SC magnetic gap,
despite the presence of AF Bragg peaks in the elastic response
\cite{YamadaPRL03,MotoyamaPRL06}. We conclude that the AF phase boundary in
fact terminates at $x_\mathrm{AF}=0.134(4)$, and that magnetic Bragg peaks
observed at higher cerium concentrations originate from regions of the
samples which were not fully oxygen-annealed. While a relatively small
volume fraction of such macroscopic remnants of the AF as-grown state can
give rise to significant Bragg scattering, our inelastic measurements are
fortunately insensitive to their presence. This conclusion is consistent
with the observation that the N\'{e}el transition is very broad in SC
samples (Fig.~\ref{stiffness}b), and also with muon spin-resonance results
\cite{UefujiPhysicaC01}, which show a significant decrease of the AF volume
fraction near $x=0.14$. We note that the spurious elastic signal from the
remnant AF regions should not be confused with the spurious elastic signal
in a magnetic field \cite{MangPRB04} due to the paramagnetic decomposition
product (Nd,Ce)$_2$O$_3$.

For as-grown NCCO \cite{MangPRL04} and for La$_2$Cu$_{1-z}$(Zn,Mg)$_z$O$_4$
\cite{VajkScience02} it was found that, to a very good approximation, the
N\'eel temperature $T_\mathrm{N}(x)$ is a contour of constant 2D correlation
length with $\xi/a=$ 200-400 and 100, respectively. Following the
observations for as-grown NCCO, we plot the extrapolated contour of
$\xi/a=400$ as a dashed curve in Fig.~\ref{phasediagram}. This estimate of
the underlying bulk N\'{e}el temperature coincides with the measured
$T_\mathrm{N}$ at $x=0.038$, but it lies systematically lower at higher
cerium concentrations, approaching $T_\mathrm{N}=0$ at
$x_\mathrm{AF}\approx0.134$.

The decrease to zero of the spin stiffness at $x_\mathrm{AF} \approx 0.134$
and the finite values of $\xi_0$ for $x>x_\mathrm{AF}$ indicates a
fundamental change in the nature of the magnetic ground state. The
contribution of the AF remnants may lead to a slight over-estimate of the
spin correlations, and consequently of $\rho_s$ and $\xi_0$, but we
emphasize that the qualitative change in behaviour is a robust result. The
NCCO phase diagram resembles those of other unconventional superconductors,
such as the heavy-fermion compound CeRhIn$_5$, in which the AF and SC phases
are believed to be separated by a first-order boundary \cite{ParkNature06}.
Although we cannot rule out a genuine underlying coexistence between AF and
SC order, such coexistence would be confined to a rather narrow doping
range. However, the behaviour of $\rho_s(x)$, which decreases continuously
by more than an order of magnitude with doping, together with the crossover
to power-law behaviour of $\xi(x,T)$ near $x=0.134$, suggests another
scenario: a second-order quantum phase transition between the AF and SC
phases. This quantum phase transition would be described by a dynamic
critical exponent of $z = 1/\nu_T \approx 1.0(5)$, which differs from the
value $z=2$ predicted for a transition from the antiferromagnet to a non-SC
paramagnet \cite{KyungPRL04}. If hyperscaling holds, the spin stiffness for
a 2D system is expected to decrease as $\rho_s \propto (x_\mathrm{AF} -
x)^{\nu_0 z}$, where $\nu_0$ is the exponent describing the divergence of
$\xi_0$ as $x_\mathrm{AF}$ is approached from above. From the approximately
linear behaviour of the spin stiffness, we therefore have $\nu_0 \approx 1$.
We cannot independently determine $\nu_0$, since we do not have sufficient
information for $\xi_0(x)$. It is also possible that the system lies above
the upper critical dimension, in which case mean-field behaviour with
$\rho_s \propto (x_\mathrm{AF} - x)^{2\beta_\mathrm{mean\ field}}$ is
expected. $\beta_\mathrm{mean\ field} = 1/2$, so this is consistent with the
observed behaviour.

Our results for $\xi(x,T)$ also have important consequences for the
relationship between AF correlations and the pseudogap physics in the
electron-doped copper oxides, which appears to be different from that of the
hole-doped materials \cite{OnosePRB04,KyungPRL04,MarkiewiczPRB04}.  The
pseudogap (charge anomalies associated with the opening of a partial gap
along the Fermi surface) has been discerned in photoemission
\cite{ArmitagePRL02,MatsuiPRL05a}, optical spectroscopy
\cite{ZimmersEPL05,OnosePRB04}, and charge transport \cite{OnosePRB04}
experiments on NCCO crystals and Pr$_{2-x}$Ce$_x$CuO$_{4\pm\delta}$ films up
to $x=0.15$ \cite{ArmitagePRL02,ZimmersEPL05}. First, our finding that the
AF phase terminates at $x_\mathrm{AF}=0.134(4)$ refutes statements that a
possible quantum phase transition at $x\approx 0.17$ is related to the
disappearance of AF order \cite{DaganPRL05,ZimmersEPL05}. Second, we find
that the spin-correlation length changes smoothly across the pseudogap
temperature $T^*$ and, remarkably, up to $x=0.145$ it follows the simple
relationship:
\begin{equation}
\frac{\xi^*}{a} = \frac{C}{x_c - x} \label{xistar}
\end{equation}
with fitting parameters $C=0.96(12)$ and $x_c = 0.171(4)$. One
interpretation of the pseudogap is that it signifies a change in the spin
scattering of the electrons as the AF correlations exceed the carrier mean
free path \cite{OnosePRB04} ($\ell$) or the single-particle thermal de
Broglie wavelength \cite{KyungPRL04} ($\xi_\mathrm{th} = \hbar v_F / \pi k_B
T$) upon cooling. Boltzmann transport theory, which might be expected to
qualitatively hold if $\ell>a$ and $T>T^*$, applied to direct-current
resistivity data for NCCO yields $\ell(T^*) \approx 2.4a$ and $\ell(T^*)
\approx 25a$ for $x=0.125$ and $x=0.15$, respectively \cite{OnosePRB04}.
This trend is qualitatively consistent with our results up to $x=0.145$ seen
in Fig.~\ref{stiffness}c. However, given the approximately linear
relationship $T^* \propto (x^* - x)$ (Fig.~\ref{phasediagram}), the thermal
de Broglie wavelength at $T^*$ exhibits the same quantitative doping
dependence as does equation~(\ref{xistar}). Using the value $v_F = 2.2
\times 10^7\,\mathrm{cm/s}$ for the bare Fermi velocity \cite{KyungPRL04},
we find that $\xi^*=2.6(2)\,\xi_\mathrm{th}$. We conclude that $T^*$ is a
crossover temperature below which the spin correlations become longer than
the thermal de Broglie wavelength, in agreement with theoretical work
\cite{KyungPRL04}. The pseudogap phenomenon in the electron-doped copper
oxides therefore results from 2D AF spin correlations, and does not appear
to be a direct precursor to superconductivity. That $\xi^* \propto
\xi_\mathrm{th}$ is obeyed so well suggests that the contributions from the
remnant AF regions to our measured $\xi(x,T)$ are negligible.

Equation~(\ref{xistar}) would suggest that $\xi^*$ diverges at $x_c \approx
x^*$.  However, as seen from Fig.~\ref{stiffness}c, this relation breaks
down near optimal doping, presumably owing to the emergence of a new length
scale that limits the development of spin correlations. Interestingly, at
optimal doping ($x=0.15$), $\xi^* \approx \xi_0$ is comparable to the SC
coherence length \cite{WangScience03} $\xi_\mathrm{SC} = 58\,\textrm{\AA}
\approx 15a$. The spin correlations near optimal doping are still relatively
large, consistent with suggestions based on Raman scattering
\cite{BlumbergPRL02} and photoemission \cite{MatsuiPRL05b} that the
$d_{x^2-y^2}$ SC order parameter is non-monotonic due to AF fluctuations.

The new experimental results for the magnetic phase diagram and pseudogap
physics contain important implications for theories of high-$T_\mathrm{c}$
superconductivity. By avoiding spurious scattering that significantly
contaminates the elastic response, we have established from measurements of
the 2D spin correlations that genuine coexistence of AF and SC order is
essentially absent in Nd$_{2-x}$Ce$_x$CuO$_{4\pm\delta}$.  On symmetry
grounds, a possible second-order quantum phase transition between these two
types of order would seem unlikely and exotic. One scenario is that an
underlying first-order transition is rendered second-order owing to
microscopic disorder \cite{AizenmanCMP90}. Such disorder is found in most
high-$T_\mathrm{c}$ superconductors \cite{EisakiPRB04} and, in the present
case, it might be the randomness associated with the Nd-Ce substitution.
Alternatively, such a phase transition may be an example of `deconfined'
quantum criticality \cite{SenthilScience04}, a new paradigm for quantum
phase transitions. To further elucidate the nature of the transition from AF
to SC order, a detailed complementary study of the superconducting critical
properties on small samples with minimal oxygen and cerium inhomogeneities
would be desirable.

\begin{acknowledgements}
We thank N. Bontemps, S. Chakravarty, S.A. Kivelson, R.S. Markiewicz and
A.-M.S. Tremblay for discussions. The work at Stanford University was
supported by grants from the Department of Energy and the National Science
Foundation. E.M.M. acknowledges support through the NSF Graduate Fellowship
programme.
\end{acknowledgements}

\bibliography{emdm}

\begin{thebibliography}{24}
\expandafter\ifx\csname natexlab\endcsname\relax\def\natexlab#1{#1}\fi
\expandafter\ifx\csname bibnamefont\endcsname\relax
  \def\bibnamefont#1{#1}\fi
\expandafter\ifx\csname bibfnamefont\endcsname\relax
  \def\bibfnamefont#1{#1}\fi
\expandafter\ifx\csname citenamefont\endcsname\relax
  \def\citenamefont#1{#1}\fi
\expandafter\ifx\csname url\endcsname\relax
  \def\url#1{\texttt{#1}}\fi
\expandafter\ifx\csname urlprefix\endcsname\relax\def\urlprefix{URL }\fi
\providecommand{\bibinfo}[2]{#2}
\providecommand{\eprint}[2][]{\url{#2}}

\bibitem[{\citenamefont{Keimer {\it et~al.}}(1992)}]{KeimerPRB92}
\bibinfo{author}{\bibfnamefont{B.}~\bibnamefont{Keimer}} \bibnamefont{{\it
  et~al.}}, \bibinfo{journal}{Phys. Rev. B} \textbf{\bibinfo{volume}{46}},
  \bibinfo{pages}{14034} (\bibinfo{year}{1992}).

\bibitem[{\citenamefont{Matsuda {\it et~al.}}(1992)}]{MatsudaPRB92}
\bibinfo{author}{\bibfnamefont{M.}~\bibnamefont{Matsuda}} \bibnamefont{{\it
  et~al.}}, \bibinfo{journal}{Phys. Rev. B} \textbf{\bibinfo{volume}{45}},
  \bibinfo{pages}{12548} (\bibinfo{year}{1992}).

\bibitem[{\citenamefont{Takagi {\it et~al.}}(1989)\citenamefont{Takagi, Uchida
  and Tokura}}]{TakagiPRL89}
\bibinfo{author}{\bibfnamefont{H.}~\bibnamefont{Takagi}},
  \bibinfo{author}{\bibfnamefont{S.}~\bibnamefont{Uchida}} \bibnamefont{and}
  \bibinfo{author}{\bibfnamefont{Y.}~\bibnamefont{Tokura}},
  \bibinfo{journal}{Phys. Rev. Lett.} \textbf{\bibinfo{volume}{62}},
  \bibinfo{pages}{1197} (\bibinfo{year}{1989}).

\bibitem[{\citenamefont{Uefuji {\it et~al.}}(2001)}]{UefujiPhysicaC01}
\bibinfo{author}{\bibfnamefont{T.}~\bibnamefont{Uefuji}} \bibnamefont{{\it
  et~al.}}, \bibinfo{journal}{Physica C} \textbf{\bibinfo{volume}{357-360}},
  \bibinfo{pages}{208} (\bibinfo{year}{2001}).

\bibitem[{\citenamefont{Uefuji {\it et~al.}}(2002)}]{UefujiPhysicaC02}
\bibinfo{author}{\bibfnamefont{T.}~\bibnamefont{Uefuji}} \bibnamefont{{\it
  et~al.}}, \bibinfo{journal}{Physica C} \textbf{\bibinfo{volume}{378-381}},
  \bibinfo{pages}{273} (\bibinfo{year}{2002}).

\bibitem[{\citenamefont{Mang {\it et~al.}}(2004{\natexlab{a}})}]{MangPRL04}
\bibinfo{author}{\bibfnamefont{P.~K.} \bibnamefont{Mang}} \bibnamefont{{\it
  et~al.}}, \bibinfo{journal}{Phys. Rev. Lett.} \textbf{\bibinfo{volume}{93}},
  \bibinfo{pages}{027002} (\bibinfo{year}{2004}{\natexlab{a}}).

\bibitem[{\citenamefont{Senthil {\it et~al.}}(2004)}]{SenthilScience04}
\bibinfo{author}{\bibfnamefont{T.}~\bibnamefont{Senthil}} \bibnamefont{{\it
  et~al.}}, \bibinfo{journal}{Science} \textbf{\bibinfo{volume}{303}},
  \bibinfo{pages}{1490} (\bibinfo{year}{2004}).

\bibitem[{\citenamefont{Armitage {\it et~al.}}(2002)}]{ArmitagePRL02}
\bibinfo{author}{\bibfnamefont{N.~P.} \bibnamefont{Armitage}} \bibnamefont{{\it
  et~al.}}, \bibinfo{journal}{Phys. Rev. Lett.} \textbf{\bibinfo{volume}{88}},
  \bibinfo{pages}{257001} (\bibinfo{year}{2002}).

\bibitem[{\citenamefont{Onose {\it et~al.}}(2004)\citenamefont{Onose, Taguchi,
  Ishizaka and Tokura}}]{OnosePRB04}
\bibinfo{author}{\bibfnamefont{Y.}~\bibnamefont{Onose}},
  \bibinfo{author}{\bibfnamefont{Y.}~\bibnamefont{Taguchi}},
  \bibinfo{author}{\bibfnamefont{K.}~\bibnamefont{Ishizaka}} \bibnamefont{and}
  \bibinfo{author}{\bibfnamefont{Y.}~\bibnamefont{Tokura}},
  \bibinfo{journal}{Phys. Rev. B} \textbf{\bibinfo{volume}{69}},
  \bibinfo{pages}{024504} (\bibinfo{year}{2004}).

\bibitem[{\citenamefont{Matsui {\it
  et~al.}}(2005{\natexlab{a}})}]{MatsuiPRL05a}
\bibinfo{author}{\bibfnamefont{H.}~\bibnamefont{Matsui}} \bibnamefont{{\it
  et~al.}}, \bibinfo{journal}{Phys. Rev. Lett.} \textbf{\bibinfo{volume}{94}},
  \bibinfo{pages}{047005} (\bibinfo{year}{2005}{\natexlab{a}}).

\bibitem[{\citenamefont{Zimmers {\it et~al.}}(2005)}]{ZimmersEPL05}
\bibinfo{author}{\bibfnamefont{A.}~\bibnamefont{Zimmers}} \bibnamefont{{\it
  et~al.}}, \bibinfo{journal}{Europhys. Lett.} \textbf{\bibinfo{volume}{70}},
  \bibinfo{pages}{225} (\bibinfo{year}{2005}).

\bibitem[{\citenamefont{Kyung {\it et~al.}}(2004)\citenamefont{Kyung,
  Hankevych, Dar\'{e} and Tremblay}}]{KyungPRL04}
\bibinfo{author}{\bibfnamefont{B.}~\bibnamefont{Kyung}},
  \bibinfo{author}{\bibfnamefont{V.}~\bibnamefont{Hankevych}},
  \bibinfo{author}{\bibfnamefont{A.-M.} \bibnamefont{Dar\'{e}}}
  \bibnamefont{and} \bibinfo{author}{\bibfnamefont{A.-M.~S.}
  \bibnamefont{Tremblay}}, \bibinfo{journal}{Phys. Rev. Lett.}
  \textbf{\bibinfo{volume}{93}}, \bibinfo{pages}{147004}
  (\bibinfo{year}{2004}).

\bibitem[{\citenamefont{Markiewicz}(2004)}]{MarkiewiczPRB04}
\bibinfo{author}{\bibfnamefont{R.~S.} \bibnamefont{Markiewicz}},
  \bibinfo{journal}{Phys. Rev. B} \textbf{\bibinfo{volume}{70}},
  \bibinfo{pages}{174518} (\bibinfo{year}{2004}).

\bibitem[{\citenamefont{Vajk {\it et~al.}}(2002)}]{VajkScience02}
\bibinfo{author}{\bibfnamefont{O.~P.} \bibnamefont{Vajk}} \bibnamefont{{\it
  et~al.}}, \bibinfo{journal}{Science} \textbf{\bibinfo{volume}{295}},
  \bibinfo{pages}{1691} (\bibinfo{year}{2002}).

\bibitem[{\citenamefont{Yamada {\it et~al.}}(2003)}]{YamadaPRL03}
\bibinfo{author}{\bibfnamefont{K.}~\bibnamefont{Yamada}} \bibnamefont{{\it
  et~al.}}, \bibinfo{journal}{Phys. Rev. Lett.} \textbf{\bibinfo{volume}{90}},
  \bibinfo{pages}{137004} (\bibinfo{year}{2003}).

\bibitem[{\citenamefont{Motoyama {\it et~al.}}(2006)}]{MotoyamaPRL06}
\bibinfo{author}{\bibfnamefont{E.~M.} \bibnamefont{Motoyama}} \bibnamefont{{\it
  et~al.}}, \bibinfo{journal}{Phys. Rev. Lett.} \textbf{\bibinfo{volume}{96}},
  \bibinfo{pages}{137002} (\bibinfo{year}{2006}).

\bibitem[{\citenamefont{Dagan {\it et~al.}}(2005)}]{DaganPRL05}
\bibinfo{author}{\bibfnamefont{Y.}~\bibnamefont{Dagan}} \bibnamefont{{\it
  et~al.}}, \bibinfo{journal}{Phys. Rev. Lett.} \textbf{\bibinfo{volume}{94}},
  \bibinfo{pages}{057005} (\bibinfo{year}{2005}).

\bibitem[{\citenamefont{Mang {\it et~al.}}(2004{\natexlab{b}})}]{MangPRB04}
\bibinfo{author}{\bibfnamefont{P.~K.} \bibnamefont{Mang}} \bibnamefont{{\it
  et~al.}}, \bibinfo{journal}{Phys. Rev. B} \textbf{\bibinfo{volume}{70}},
  \bibinfo{pages}{094507} (\bibinfo{year}{2004}{\natexlab{b}}).

\bibitem[{\citenamefont{Park {\it et~al.}}(2006)}]{ParkNature06}
\bibinfo{author}{\bibfnamefont{T.}~\bibnamefont{Park}} \bibnamefont{{\it
  et~al.}}, \bibinfo{journal}{Nature} \textbf{\bibinfo{volume}{440}},
  \bibinfo{pages}{65} (\bibinfo{year}{2006}).

\bibitem[{\citenamefont{Wang {\it et~al.}}(2003)}]{WangScience03}
\bibinfo{author}{\bibfnamefont{Y.}~\bibnamefont{Wang}} \bibnamefont{{\it
  et~al.}}, \bibinfo{journal}{Science} \textbf{\bibinfo{volume}{299}},
  \bibinfo{pages}{86} (\bibinfo{year}{2003}).

\bibitem[{\citenamefont{Blumberg {\it et~al.}}(2002)}]{BlumbergPRL02}
\bibinfo{author}{\bibfnamefont{G.}~\bibnamefont{Blumberg}} \bibnamefont{{\it
  et~al.}}, \bibinfo{journal}{Phys. Rev. Lett.} \textbf{\bibinfo{volume}{88}},
  \bibinfo{pages}{107002} (\bibinfo{year}{2002}).

\bibitem[{\citenamefont{Matsui {\it
  et~al.}}(2005{\natexlab{b}})}]{MatsuiPRL05b}
\bibinfo{author}{\bibfnamefont{H.}~\bibnamefont{Matsui}} \bibnamefont{{\it
  et~al.}}, \bibinfo{journal}{Phys. Rev. Lett.} \textbf{\bibinfo{volume}{95}},
  \bibinfo{pages}{017003} (\bibinfo{year}{2005}{\natexlab{b}}).

\bibitem[{\citenamefont{Aizenman and Wehr}(1990)}]{AizenmanCMP90}
\bibinfo{author}{\bibfnamefont{M.}~\bibnamefont{Aizenman}} \bibnamefont{and}
  \bibinfo{author}{\bibfnamefont{J.}~\bibnamefont{Wehr}},
  \bibinfo{journal}{Commun. Math. Phys.} \textbf{\bibinfo{volume}{130}},
  \bibinfo{pages}{489} (\bibinfo{year}{1990}).

\bibitem[{\citenamefont{Eisaki {\it et~al.}}(2004)}]{EisakiPRB04}
\bibinfo{author}{\bibfnamefont{H.}~\bibnamefont{Eisaki}} \bibnamefont{{\it
  et~al.}}, \bibinfo{journal}{Phys. Rev. B} \textbf{\bibinfo{volume}{69}},
  \bibinfo{pages}{064512} (\bibinfo{year}{2004}).

\end{thebibliography}

\end{document}